**Synthesis, Crystal Structure, and Physical Properties of New Layered Oxychalcogenide $La_2O_2Bi_3AgS_6$**

Yudai Hijikata[1], Tomohiro Abe[2], Chikako Moriyoshi[2], Yoshihiro Kuroiwa[2], Yosuke Goto[1], Akira Miura[3], Kiyoharu Tadanaga[3], Yongming Wang[4], Osuke Miura[1], and Yoshikazu Mizuguchi[1]*

1. Graduate School of Science and Engineering, Tokyo Metropolitan University, 1-1 Minami-osawa, Hachioji 192-0397, Japan
2. Department of Physical Science, Hiroshima University, 1-3-1 Kagamiyama, Higashihiroshima 739-8526, Japan
3. Faculty of Engineering, Hokkaido University, Kita-13, Nishi-8, Kita-ku, Sapporo 060-8628, Japan
4. Creative Research Institution, Hokkaido University, Kita-21, Nishi-10, Kita-ku, Sapporo 001-0021, Japan

Abstract

We have synthesized a new layered oxychalcogenide $La_2O_2Bi_3AgS_6$. From synchrotron X-ray diffraction and Rietveld refinement, the crystal structure of $La_2O_2Bi_3AgS_6$ was refined using a model of the *P*4/*nmm* space group with $a = 4.0644(1)$ Å and $c = 19.412(1)$ Å, which is similar to the related compound $LaOBiPbS_3$, while the interlayer bonds (M2-S1 bonds) are apparently shorter in $La_2O_2Bi_3AgS_6$. The tunneling electron microscopy (TEM) image confirmed the lattice constant derived from Rietveld refinement ($c \sim 20$ Å). The electrical resistivity and Seebeck coefficient suggested that the electronic states of $La_2O_2Bi_3AgS_6$ are more metallic than those of $LaOBiS_2$ and $LaOBiPbS_3$. The insertion of a rock-salt-type chalcogenide into the van der Waals gap of $BiS_2$-based layered compounds, such as $LaOBiS_2$, will be a useful strategy for designing new layered functional materials in the layered chalcogenide family.

1. Introduction

Layered compounds have been extensively studied in the fields of superconductivity and thermoelectric materials [1-6] because of the high flexibility of the stacking structure and the constituent elements. In addition, low-dimensional electronic states can be generated, and thus, unconventional superconductivity or exotic functionality has been revealed in several materials.

Recently, $BiS_2$-based layered compounds, whose crystal structure is composed of alternate stacks of electrically conducting $BiS_2$ bilayers and various kinds of electrically insulating (blocking) layers, have drawn much attention as superconductors [7-16] and thermoelectric materials [17-21]. The typical parent phase is $REOBiS_2$ (RE: Rare earth element). From band calculations [22,23], the parent phase is basically an insulator with a band gap. The conduction band just above the Fermi energy is mainly composed of Bi-6p ($p_x$ and $p_y$) bands. Typically, the partial substitution of $O^{2-}$ at the blocking REO layers with $F^-$ generates electron carriers in the $BiS_2$ layer, and superconductivity has been observed in $REO_{1-x}F_xBiS_2$ [9-14]. The highest reported superconducting transition temperature ($T_c$) in this family is 11 K for $LaO_{0.5}F_{0.5}BiS_2$ [24], although a possible high $T_c$ phase with $T_c \sim 20$ K was detected for Nd(O,F)$BiS_2$ single crystal using scanning tunneling spectroscopy [25]. In addition, the possibility of unconventional mechanisms of the superconductivity in $BiS_2$-based compounds has been recently proposed [26-29]. Thus, further exploration of the new superconducting phases and clarification of the mechanisms are desired. The $BiS_2$-based compounds exhibit high thermoelectric performance. The highest reported thermoelectric dimensionless figure of merit ($ZT$) in this family is 0.36 at 650 K for LaOBiSSe [18]. The high $ZT$ was achieved by the enhancement of carrier mobility and the low thermal conductivity [18,19]. To further enhance $ZT$, exploration of new $BiS_2$-type layered compounds is needed.

Very recently, we have reported the crystal structure and site selectivity of $LaOBiPbS_3$ [30], which was discovered to be a thermoelectric material [20]. From neutron diffraction, we revealed the site selectivity of Bi and Pb in $LaOBiPbS_3$. On the basis of crystal structure refinements with neutron and synchrotron X-ray diffraction, it was found that the crystal structure of $LaOBiPbS_3$ can be regarded as the alternate stacks of PbS layers and $LaOBiS_2$ layers. In addition, the insertion of PbS layers results in an electronic structure (at around the Fermi energy) apparently different from that of $LaOBiS_2$. If another type of chalcogenide layer can be sandwiched by $LaOBiS_2$ layers, a new layered oxychalcogenide can be synthesized with a modified electronic structure, which may be useful for the emergence of superconductivity or high thermoelectric performance. In this study, we have designed and synthesized a new layered bismuth oxysulfide $La_2O_2Bi_3AgS_6$ with the concept obtained from the crystal structure of $LaOBiPbS_3$. The crystal structure of $La_2O_2Bi_3AgS_6$ is composed of stacks of $(Bi_{0.66}Ag_{0.34})$S chalcogenide layers and $LaO(Bi_{0.9}Ag_{0.1})S_2$ layers, as displayed in Fig. 1.

2. Experimental Details

La$_2$O$_2$Bi$_3$AgS$_6$ polycrystalline samples were prepared by a solid-state reaction method. Powders of Bi$_2$O$_3$ (99.9%), La$_2$S$_3$ (99.9%), and Ag (99.9%) and grains of Bi (99.999%) and S (99.99%) with a nominal composition of La$_2$O$_2$Bi$_3$AgS$_6$ were mixed in a mortar, pelletized, sealed in an evacuated quartz tube, and heated at 720 °C for 15 h. The obtained sample was ground, mixed to homogenize it, pelletized, and heated at 720 °C for 15 h. The phase purity of the prepared samples and the optimal annealing conditions were examined using laboratory X-ray diffraction (XRD) with Cu-$K\alpha$ radiation. Synchrotron XRD was performed at BL02B2 of SPring-8 with an energy of 25 keV (project number: 2016B0074). The synchrotron XRD experiments were performed with a sample rotator system at room temperature, and the diffraction data were collected using a high-resolution one-dimensional semiconductor detector (MYTHEN) [31] with a step of $2\theta = 0.006°$. The crystal structure parameters were refined using the Rietveld method with RIETAN-FP [32]. Schematic images of the crystal structure were drawn using VESTA [33]. The ratio of Bi to Ag in the obtained sample was examined using energy dispersive X-ray spectroscopy (EDX) with a TM3030 microscope. Tunneling electron microscopy (TEM; JEOL, JEM-2010) was performed with an accelerating voltage of 200 keV. An ethanol suspension of the powder was dispersed on a commercial grid. The temperature dependence of the electrical resistivity was measured by the four-probe method. For high-temperature measurements, the probes were mechanically attached to the sample, while Ag paste was used to attach Au wires to the sample for low-temperature measurements. The Seebeck coefficient was measured by the four-probe method as well as high-temperature resistivity. In this article, the sample is named La$_2$O$_2$Bi$_3$AgS$_6$, which is the starting nominal composition.

3. Results and Discussion

As shown in Fig. 1(c), the refined structure for La$_2$O$_2$Bi$_3$AgS$_6$ was found to be similar to that of LaOBiPbS$_3$ with the space group of *P*4/*nmm* [20,30] [Fig. 1(b)]. La$_2$O$_2$S (*P*-3*m*1: #164) and AgBiS$_2$ (*Fm*-3*m*: #225) impurity phases were also detected. The LaOBiS$_2$ phase with the space group of *P*4/*nmm* was hardly detected. Although the peaks of AgBiS$_2$ are broad, the existence of AgBiS$_2$ was also confirmed by the observation of particles with the composition of Ag:Bi:S = 1:1:2 by EDX. Figure 2 shows the multiphase Rietveld refinement profile obtained using synchrotron XRD. With the *P*4/*nmm* model and two impurity phases, the XRD pattern was refined with a resulting reliability factor of $R_{wp} = 11.6\%$. $R_{wp}$ was not as low as that obtained for LaOBiPbS$_3$ ($R_{wp} = 7.5\%$ [30]), measured in the same beamline and analyzed with the same structural model. This is attributed to the shoulder of the peaks at a lower angle than the major phase, which may be due to the low crystallinity of the phase. The obtained structure parameters are listed in Table I. The lattice constants of $a = 4.0644(1)$ Å and $c = 19.412(1)$ Å are smaller than those of LaOBiPbS$_3$: $a = 4.09716(4)$ Å and $c = 19.7933(2)$ Å. The differences in lattice constants are consistent with the smaller ionic radius of Ag$^+$ than that of Pb$^{2+}$: 1.15 Å for Ag$^+$ and 1.19 Å for Pb$^{2+}$ when the coordinate number is 6. In addition, the smaller lattice constants are consistent with the large amount of Bi$^{3+}$ ions, with an ionic radius of 1.03 Å, in La$_2$O$_2$Bi$_3$AgS$_6$.

In Fig. 1, Bi-S bonding is depicted when the bond length is shorter than 2.95 Å. The M2-S1 bond length [2.82(3) Å] is clearly shorter than that in LaOBiPbS$_3$ [3.018(9) Å], which should enhance the interlayer transfer of carriers. The short interlayer bonding may be due to the suppression of the effects of lone pairs because Bi$^{3+}$ and Pb$^{2+}$ have 6s lone pairs, which affect the local structure of the conducting layers in LaOBiS$_2$ and LaOBiPbS$_3$. Because Ag$^+$ does not have lone pairs, the interlayer bonding may decrease, as observed in the InS$_2$-based compound LaOInS$_2$ [34]. From the refinement, the Bi concentration can be estimated as 3.1(1), which is almost the same as the starting nominal composition. EDX analysis also confirmed that the composition of most of the area almost corresponded to the starting nominal composition.

Figure 3 shows a TEM image of the La$_2$O$_2$Bi$_3$AgS$_6$ sample. We confirmed the existence of fringes with a length of approximately 20 Å, corresponding to the lattice constant of $c$ obtained from the Rietveld refinement. Nonetheless, fringes with a length of 14 Å, close to the $c$-axis of LaOBiS$_2$, were seen [9]. Such disorder has also been seen in Bi$_4$O$_4$S$_3$ and related compounds [35,36]. Although clear diffraction peaks of LaOBiS$_2$ were not seen in the XRD pattern, TEM diffraction suggests that the exact structure of La$_2$O$_2$Bi$_3$AgS$_6$ is rather complicated, and the above Rietveld analysis can thus derive the average structure without considering the local disorder.

Figure 4 shows the temperature dependences of the electrical resistivity for LaOBiS$_2$ [Fig. 1(a)], LaOBiPbS$_3$ [Fig. 1(b)] [37] and La$_2$O$_2$Bi$_3$AgS$_6$ [Fig. 1(c)]. As mentioned in the introduction, LaOBiS$_2$ becomes a superconductor when electron carriers are doped by elemental substitution. Notably, the resistivity of La$_2$O$_2$Bi$_3$AgS$_6$ is clearly lower than that of LaOBiS$_2$ over a wide temperature range. In addition, the increase in resistivity at low temperatures observed for LaOBiS$_2$ is not observed for La$_2$O$_2$Bi$_3$AgS$_6$, and almost temperature-independent resistivity is observed for La$_2$O$_2$Bi$_3$AgS$_6$. Figure 5 shows the temperature dependences of the Seebeck coefficient for LaOBiS$_2$, LaOBiPbS$_3$, and La$_2$O$_2$Bi$_3$AgS$_6$. The absolute value of the Seebeck coefficient is the smallest for La$_2$O$_2$Bi$_3$AgS$_6$ among these compounds. On the basis of the electrical resistivity and Seebeck coefficient, the carrier concentration in La$_2$O$_2$Bi$_3$AgS$_6$ is higher than that in LaOBiPbS$_3$ and LaOBiS$_2$. For LaOBiPbS$_3$, band calculations revealed that the band structure is markedly changed by the insertion of a PbS layer into the van der Waals gap of LaOBiS$_2$. The valence and conduction bands of LaOBiS$_2$ and related BiS$_2$-based compounds are basically composed of Bi-6p and S-3p orbitals in the same BiS plane [22,23]. In contrast, the valence band top of LaOBiPbS$_3$ is mainly composed of S-3p bands of the PbS block while the conduction band bottom is mainly composed of Bi-6p bands [30]. A similar situation, a valence band bottom composed of S bands of a NaCl-type Ag-rich layer, can also be expected in $La_2O_2Bi_3AgS_6$, and the carrier excitation may be enhanced in the present Ag case. The enhanced interlayer bonding (M2-S1 bonding) mentioned in the structure analysis part may be related to the enhanced electrical conductivity and increased carrier concentration.

In the present study, we did not observe a superconducting transition. From magnetization measurements using a superconducting quantum interference device (SQUID) magnetometer for $La_2O_2Bi_3AgS_6$, we confirmed the absence of a superconducting transition down to 2 K. In addition, from the resistivity measurements down to 3.8 K, we did not observe any onset of a superconducting transition. We expect that $La_2O_2Bi_3AgS_6$ will become a superconductor upon carrier doping or structural optimization, as observed for other $BiS_2$-based superconductors [38].

4. Conclusion

We have reported the synthesis of a new layered oxychalcogenide $La_2O_2Bi_3AgS_6$. On the basis of the material design concept revealed in $LaOBiPbS_3$, rock-salt-type $(Ag_{0.66}Bi_{0.34})S$ layers are inserted into the $BiS_2$-based $LaO(Bi_{0.9}Ag_{0.1})S_2$. From synchrotron XRD, Rietveld refinement, and TEM, the crystal structure of $La_2O_2Bi_3AgS_6$ was revealed, as shown in Fig. 1(c), which is similar to that of $LaOBiPbS_3$ with the *P*4/*nmm* space group, and the lattice constant estimated from the Rietveld refinement and TEM had similar values ($c \sim 20$ Å). The temperature dependences of the electrical resistivity and Seebeck coefficient suggested that the electronic states of $La_2O_2Bi_3AgS_6$ are more metallic than those of $LaOBiS_2$ and $LaOBiPbS_3$, which may be due to the formation of short M2-S1 bonds along the *c*-axis. The insertion of a rock-salt-type chalcogenide into the van der Waals gap of $BiS_2$-based layered compounds, such as $LaOBiS_2$, will be a useful strategy for designing new layered functional materials in the layered chalcogenide family.

Acknowledgements

This work was partly supported by Grants-in-Aid for Scientific Research (Nos. 15H05886, 16H04493, and 17H04950) and JST-CREST (No. JPMJCR16Q6), Japan. The SXRD experiments were performed under Proposal No. 2016B0074 at SPring-8. The TEM analysis was carried out with a JEM-2010 at the Joint-use Facilities: Laboratory of Nano-Micro Material Analysis, Hokkaido University, supported by the Material Analysis and Structure Analysis Open Unit (MASAOU).

Table I. Crystal structure parameters of the $La_2O_2Bi_3AgS_6$ sample.

| Nominal composition | $La_2O_2Bi_3AgS_6$ |
|---|---|
| Space group | $P4/nmm$ (#129) |
| $a$ (Å) | 4.0644(1) |
| $c$ (Å) | 19.412(1) |
| $V$ (Å$^3$) | 320.68(2) |
| La: (x, y, z) / $U$ (Å$^2$) | (0, 1/2, 0.0606(3)) / 0.005(2) |
| O: (x, y, z) / $U$ (Å$^2$) | (0, 0, 0) / 0.013(fixed) |
| M1: (x, y, z) / $U$ (Å$^2$) | (1/2, 0, 0.5705(3)) / 0.023(2) |
| M2: (x, y, z) / $U$ (Å$^2$) | (1/2, 0, 0.2649(2)) / 0.007(1) |
| S1: (x, y, z) / $U$ (Å$^2$) | (1/2, 0, 0.716(1)) / 0.013(fixed) |
| S2: (x, y, z) / $U$ (Å$^2$) | (1/2, 0, 0.145(1)) / 0.013(fixed) |
| S3: (x, y, z) / $U$ (Å$^2$) | (1/2, 0, 0.423(1)) / 0.013(fixed) |
| Occupancy at M1 | Bi : Ag = 0.90(3) : 0.10(3) |
| Occupancy at M2 | Bi : Ag = 0.66(4) : 0.34(3) |
| M1-S1 distance (Å) | 2.898(3) |
| M1-S2 distance (Å) | 2.33(2) |
| M1-S3 distance (Å) | 3.07(2) |
| M2-S1 distance (Å) | 2.82(3) |
| M2-S3 distance (Å) (along $a$-axis) | 2.8767(9) |
| M2-S3 distance (Å) (along $c$-axis) | 2.86(3) |
| $R_{wp}$ (%) | 11.6 |
| $La_2O_2S$ impurity (%) | 6 |
| $AgBiS_2$ impurity (%) | 6 |

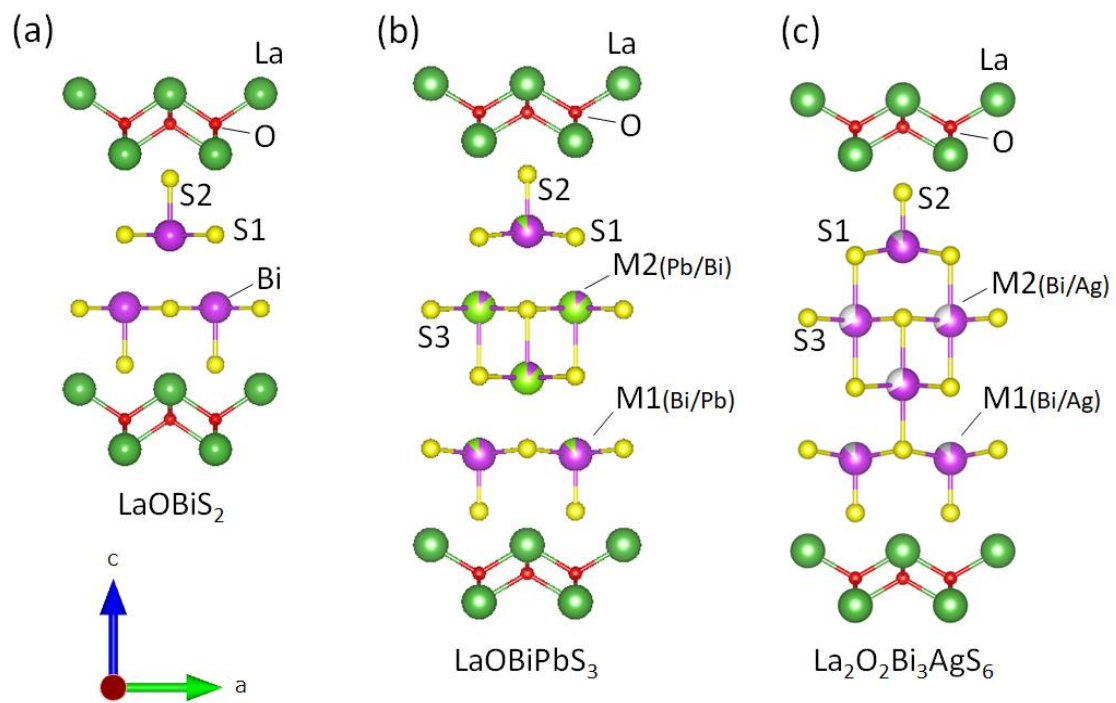

Fig. 1. (Color Online) Schematic images of the crystal structure of LaOBiS$_2$, LaOBiPbS$_3$, and La$_2$O$_2$Bi$_3$AgS$_6$. Bi-S bonds are displayed when the Bi-S distance is shorter than 2.95 Å.

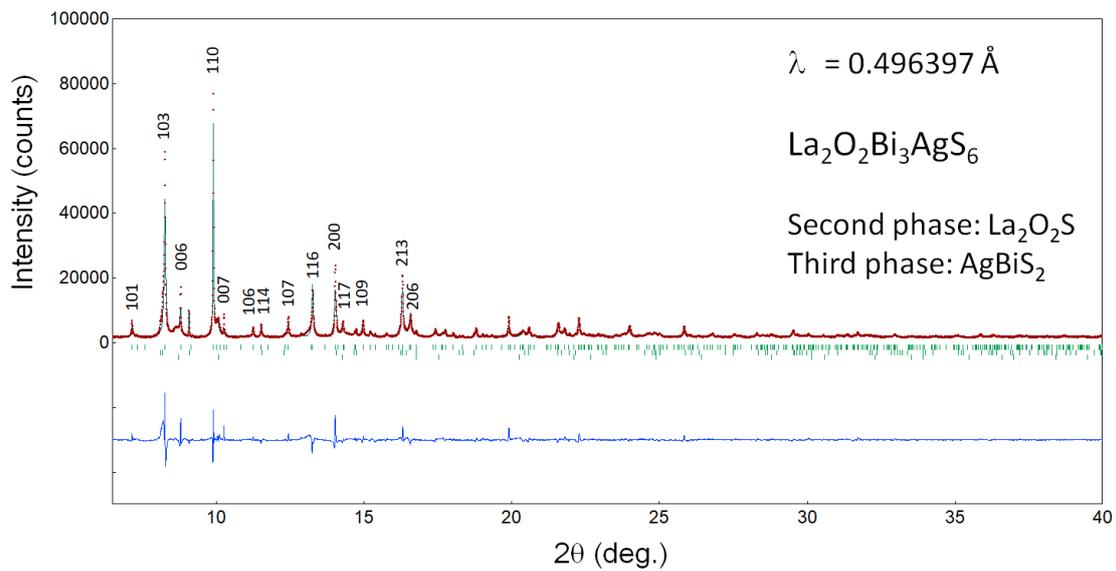

Fig. 2. (Color Online) Synchrotron XRD pattern and the Rietveld fitting for $La_2O_2Bi_3AgS_6$. The Rietveld refinement was performed with two impurity phases of $La_2O_2S$ and $AgBiS_2$. The numbers on XRD peaks are Miller indices.

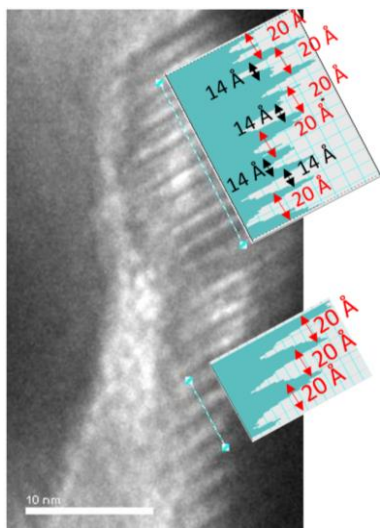

Fig. 3. (Color Online) TEM image of $La_2O_2Bi_3AgS_6$. The scale bar represents 10 nm.

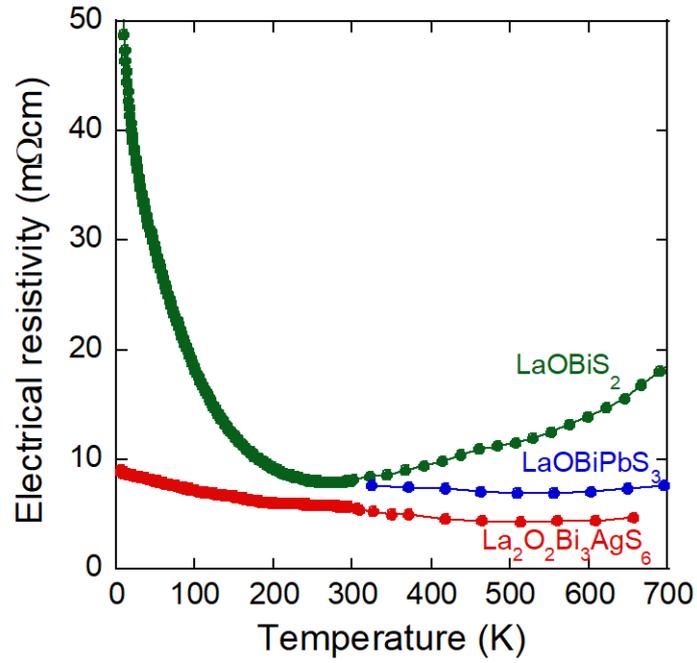

Fig. 4. (Color Online) Temperature dependences of electrical resistivity for LaOBiS$_2$ [37], LaOBiPbS$_3$ [37], and La$_2$O$_2$Bi$_3$AgS$_6$.

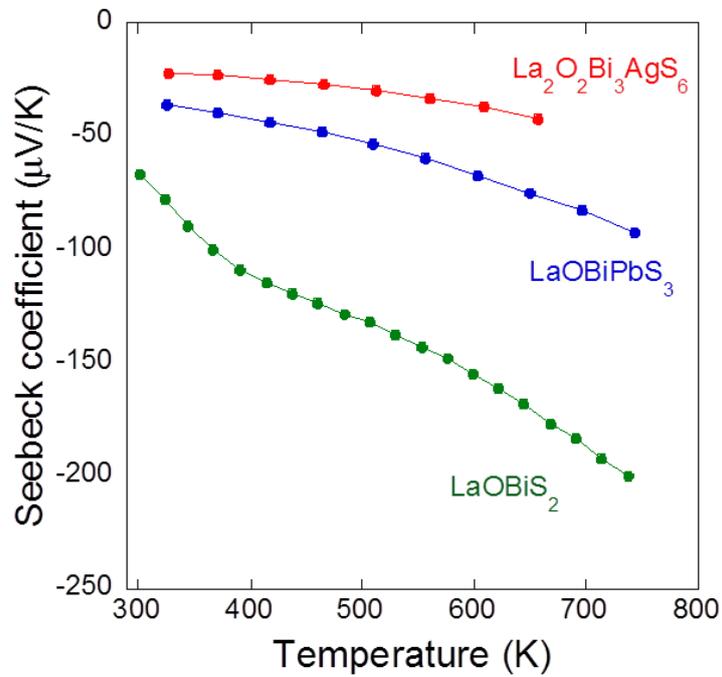

Fig. 5. (Color Online) Temperature dependences of Seebeck coefficient for LaOBiS$_2$ [37], LaOBiPbS$_3$ [37], and La$_2$O$_2$Bi$_3$AgS$_6$.